# Uncertainty-Aware Graph Neural Reconstruction of Urban Temperature Fields from Sparse Sensors under Deployment Constraints

Reda Snaiki[1]; Abdelatif Merabtine[1]

[1] *Department of Construction Engineering, École de Technologie Supérieure, Université du Québec, Montréal, Québec, Canada*

**Abstract:** Reconstructing spatially continuous daily temperature fields from sparse observations is important for urban climate monitoring and heat-risk analysis, but practical deployments are limited by sensor budgets and spacing constraints. This study proposes an uncertainty-aware graph neural network (GNN) framework for reconstructing daily maximum temperature fields from sparse sensors while supporting distance-constrained sensor placement and probabilistic exceedance mapping. The model predicts both the temperature field and a spatially varying predictive uncertainty field using a graph-attention-based mean-residual architecture trained with a Gaussian negative log-likelihood. Sensor placement is addressed using a Proper Orthogonal Decomposition with QR factorization (POD-QR) strategy with a 4 km minimum inter-sensor distance constraint and is compared with random feasible placement and farthest-point sampling. The framework is evaluated over a Montréal-area polygon using Daymet v4.1 daily temperature data (1 km resolution) under a strict temporal hold-out protocol (training: 2020–2023; testing: 2024). Across sensor budgets (10–40 sensors), the proposed GNN consistently outperforms inverse distance weighting and ordinary kriging in RMSE and MAE on unobserved nodes. Sensor-placement effects are most pronounced at low budgets and diminish at higher budgets, with a practical saturation regime emerging around 30 sensors under the imposed spacing constraint. Probabilistic evaluation further shows improved uncertainty calibration with increasing sensor density and a better sharpness-calibration trade-off than kriging. These results support the proposed framework as an effective tool for uncertainty-aware temperature field reconstruction and decision-oriented heat-risk mapping.

**Keywords:** Graph neural networks; temperature reconstruction; sparse sensors; uncertainty quantification; sensor placement.

## 1. Introduction

High-resolution temperature fields are increasingly needed for urban climate monitoring, heat-risk assessment, and operational decision-making. Applications such as heat exposure mapping, hotspot identification, targeted mitigation planning, and threshold-based alerting require spatially continuous information, while direct observations are usually available only at a limited number of stations or sensors. This creates a fundamental mismatch between decision needs and measurement reality: practitioners need dense temperature maps, but monitoring systems typically provide sparse observations. The challenge is further compounded by practical deployment constraints. Sensor locations are often restricted by accessibility, maintenance logistics, communications availability, security, and spacing requirements intended to avoid redundant collocated measurements. As a result, the relevant problem in practice is not interpolation from an idealized network, but reconstruction of full temperature fields from sparse and constrained sensing configurations. Key challenges include data sparsity as limited sensor coverage leads to poor reconstruction accuracy and robustness; urban heterogeneity as a result of variability in land use and building morphology; uncertainty quantification; sensor quality and their deployment strategy (X. Liu & Ren, 2026; Yu et al., 2024).

Classical spatial interpolation methods remain widely used because they are interpretable and relatively easy to implement. Inverse distance weighting is attractive for its simplicity, but it relies on a fixed distance-based weighting rule and does not learn spatial structure from data (Chen & Liu, 2012; Fan et al., 2016; Y. Liu et al., 2025; Ozelkan et al., 2013; Shahidi & Abedini, 2018; Yan et al., 2021; Zhu et al., 2022). This can limit its ability to represent complex temperature patterns beyond smooth geometric decay, and it does not natively provide a principled probabilistic uncertainty estimate. Kriging offers a stronger statistical framework and can provide both a predicted mean and an associated prediction variance, making it a natural benchmark for uncertainty-aware interpolation. However, kriging performance depends strongly on modeling assumptions, including variogram specification and parameter estimation, and its uncertainty quality is tied to those assumptions (Qiao et al., 2018). In heterogeneous urban and peri-urban environments, departures from stationarity or other idealized covariance assumptions can degrade performance. In addition, repeated kriging analyses across many days, sensor budgets, and placement scenarios can become computationally burdensome in comparative studies.

Recent advances in machine learning have demonstrated significant potential as alternatives to traditional numerical approaches for spatial field reconstruction from sparse observations. Bilevel sparse least-squares Support Vector Machines (SVMs) and U-Net architectures enhance reconstruction accuracy and robustness without requiring dense partial differential equation (PDE) solutions or large training datasets (Delgado-Enales et al., 2025). Physics-Informed Neural Networks (PINNs) integrate sparse sensor measurements with governing physical equations to reconstruct temperature and velocity fields, thereby bridging the gap between direct observations and purely data-driven regression models (Lopez et al., 2026). Low-rank tensor completion methods effectively capture both global and local spatiotemporal patterns, enabling accurate reconstruction of dynamic urban states from incomplete data (Wang et al., 2023). Furthermore, hybrid deep learning architectures, including CNN-LSTM, BiGRU, and other combined models, consistently outperform standalone approaches by jointly modeling spatial and temporal dependencies in urban climate data (Gu et al., 2024; Khala et al., 2025; Vallileka et al., 2025; Yasavoli et al., 2025).

Incorporating temporal dynamics via spatiotemporal graph models is critical for accurate and robust urban temperature field reconstruction from sparse time-varying data (Abdullah Al Mamun et al., 2018; Ding et al., 2026; H. He et al., 2026; Inagaki et al., 2022; Zhang et al., 2024). Spatiotemporal graph models encode both spatial and temporal relationships, enabling dynamic reconstruction from sparse sensors (Deuri et al., 2025). Graph neural networks (GNNs) are well suited because they represent spatial domains as graphs, support local message passing, and can incorporate positional information while handling irregular masks and domain geometries. For temperature mapping, this offers a flexible way to learn nonlinear relationships between sparse measurements and dense spatial fields, potentially outperforming fixed-form interpolation methods when spatial dependencies are complex or context-dependent (Delgado-Enales et al., 2025). GNNs with attention, diffusion convolution, and spatiotemporal modules are highly effective for urban temperature field modeling, especially when integrated with uncertainty quantification (Feng et al., 2023; Fu et al., 2022; M. He et al., 2025; Hu et al., 2024; Just et al., 2025; H. Li et al., 2026; J. Li et al., 2025; Oláh et al., 2025; Sengupta et al., 2026; Yu et al., 2024).

However, important gaps remain in current learning-based interpolation approaches. As for hybrid models that integrate remote sensing, 3D urban morphology, and physical process knowledge for high-resolution temperature estimation (Klimenka et al., 2025; Patel et al., 2023; Zang et al., 2023), many methods emphasize deterministic accuracy and provide only point estimates, without calibrated predictive uncertainty. This limits their usefulness in risk-oriented applications where confidence-aware outputs are essential. In addition, sensor placement and field reconstruction are often treated as separate problems, with model performance evaluated for assumed sensor locations but without systematically addressing how placement design under realistic constraints affects reconstruction and uncertainty.

While GNNs improve predictive accuracy by explicitly modeling relational dependencies, their deployment in climate-sensitive and decision-critical applications necessitates robust uncertainty quantification. Two primary sources of uncertainty are typically distinguished in this context: Aleatoric uncertainty, which arises from inherent variability and noise in environmental observations and processes (e.g., sensor noise, microclimatic fluctuations); and Epistemic uncertainty, which reflects model uncertainty due to limited data coverage, incomplete process representation, or parameter uncertainty, and which is reducible with additional information. Within GNN-based urban temperature modeling, deep ensembles and Bayesian learning frameworks have emerged as leading approaches for disentangling and quantifying these uncertainty components.

Deep ensemble methods approximate epistemic uncertainty by training multiple independently initialized GNNs and aggregating their predictive distributions (Gustafsson et al., 2020; Scalia et al., 2020). Variability across ensemble members captures model uncertainty, while each model can additionally be parameterized to estimate aleatoric uncertainty through heteroscedastic output distributions. Deep ensembles are widely adopted due to their conceptual simplicity, strong empirical performance, and compatibility with existing GNN architectures.

Bayesian frameworks, including Bayesian neural networks and variational inference applied to GNNs, provide a theoretically grounded formulation in which epistemic uncertainty is represented through posterior distributions over model parameters (Munikoti et al., 2023). In urban temperature applications, Bayesian GNNs enable principled uncertainty propagation across graph structures and temporal dependencies, thereby supporting risk-aware forecasting and spatial generalization assessment. However, their practical adoption is sometimes constrained by computational cost and optimization complexity, especially for large urban graphs.

Graph Attention Networks (GATs) improve inference accuracy for environmental sensor data by selectively modeling and decoupling short- and long-term relational patterns. In their study, Hu et al. (Hu et al., 2024) developed DualSTN model, a spatiotemporal inference framework designed to effectively capture both short- and long-term dynamics for sparse sensor stations monitoring air quality in Beijing. To this end, the model was decomposed into two complementary modules: JST-GAT and SG-GRU. The JST-GAT component was dedicated to learning fine-grained joint spatiotemporal dependencies over short horizons, whereas the SG-GRU module leverages a time-skipping strategy to model long-term temporal patterns. Zhang (Zhang, 2025) noted that, even though some studies incorporate physical laws into GAT (e.g. wind speed prediction), this approach is underexplored for urban temperature fields, where physical consistency is critical. Tong et al. (Tong et al., 2022) and Haghbayan et al. (Haghbayan et al., 2025) have, respectively, applied GAT models for mobile crowd sensing data reconstruction and for modeling $PM_{2.5}$ concentrations in urban environments and stated that environment parameters (eg. Air temperature) exhibit complex spatio-temporal correlations, which are challenging to model accurately. Moreover, sparse and noisy urban sensing data, due to limitations in sensor coverage and environmental factors, pose significant challenges for accurate reconstruction (Mo et al., 2021).

Existing GAT models often struggle with error propagation and capturing high-order dependencies. Gao et al. (Gao et al., 2025) generated an intelligent 3D temperature field for cryogenic target surfaces, integrating a GAT-Transformer Network (GTN) with a BiLSTM network to intelligently compute time-series prediction data of 3D temperature fields that adhere to physical laws. They demonstrated that The GTN effectively simulates theoretical infrared temperature fields under cryogenic conditions, producing realistic thermal variations to support high-quality training data for target recognition. However, high-resolution urban temperature modeling requires scalable GAT architectures since current models often face computational inefficiencies when applied to large-scale urban datasets.

This study addresses these gaps by developing an uncertainty-aware graph neural network framework for daily temperature field reconstruction from sparse sensor observations under practical spacing constraints. The proposed approach predicts both a reconstructed temperature field and a spatially varying predictive uncertainty field, enabling accuracy assessment and probabilistic interpretation within the same framework. The evaluation is conducted under a strict temporal generalization protocol, with training performed on 2020–2023 data and testing on 2024 data, and includes both deterministic and probabilistic metrics such as RMSE, MAE, CRPS, NLL, and empirical coverage. The study also examines distance-constrained sensor placement using a Proper Orthogonal Decomposition combined with QR factorization (POD-QR) within a 4 km minimum inter-sensor spacing requirement, and compares it with random feasible placement and farthest-point sampling to quantify how placement design affects downstream reconstruction quality. Finally, the predicted uncertainty is translated into exceedance probability maps to demonstrate how the framework can support decision-oriented temperature-risk applications. The remainder of the paper presents the methodology, application results, discussion, and conclusions in that order.

## 2. Methodology

### 2.1 Problem statement and objective

Urban-scale temperature monitoring is inherently sparse because only a limited number of environment sensors can be deployed in practice. However, many applications, including heat exposure mapping and operational heat-risk assessment, require spatially continuous temperature fields at fine resolution. Let $y_t(x)$ denote the daily near-surface air temperature (here, daily maximum temperature $t_{max}$) at day $t$ and spatial location $x$ over a discretized domain. After masking invalid pixels (e.g., outside the study polygon or over missing grid cells), the domain is represented by $N$ valid grid nodes $\{x_i\}_{i=1}^{N}$, and the temperature field at day $t$ is written as a vector:

$$\boldsymbol{y}_t = \left[y_{t,1}, y_{t,2}, \dots, y_{t,N}\right]^T \in \mathbb{R}^N \tag{1}$$

where $y_{t,i}$ is the temperature at node $i$ (in °C). On each day $t$, only a subset of nodes is observed through a sparse sensor network. Let $S \subset \{1, \dots, N\}$ be the index set of sensor locations (with $|S| = m$), and let $\boldsymbol{y}_{t,S}$ denote the corresponding observed temperatures. The remaining nodes $\bar{S} = \{1, \dots, N\} \backslash S$ are unobserved and must be inferred. Our goal is therefore to learn a spatial reconstruction operator that maps sparse observations to a full-field estimate:

$$\left(\boldsymbol{y}_{t,S}, S\right) \mapsto \widehat{\boldsymbol{\mu}}_t \in \mathbb{R}^N \tag{2}$$

where $\widehat{\boldsymbol{\mu}}_t$ is the predicted mean temperature field. Beyond point estimates, decision-making under sparse observations requires a quantified notion of confidence. We therefore seek a second output that characterizes predictive uncertainty at each grid node. Specifically, the model predicts a heteroscedastic (spatially varying) uncertainty field, reported in physical units as the predictive standard deviation associated with the reconstructed mean field $\widehat{\boldsymbol{\mu}}_t$.:

$$\widehat{\boldsymbol{\sigma}}_t = \left[\hat{\sigma}_{t,1}, \hat{\sigma}_{t,2}, \dots, \hat{\sigma}_{t,N}\right]^T \in \mathbb{R}_+^N \tag{3}$$

This uncertainty is used to produce decision-oriented products such as exceedance probability maps. Finally, because the sensor locations $S$ are not arbitrary in practice, a companion design problem is considered to select informative sensor sets under operational constraints (e.g., minimum inter-sensor distance).

### 2.2 Proposed framework overview

This section provides a high-level description of the proposed framework for daily temperature field reconstruction with uncertainty quantification from sparse observations. The framework is designed for the problem introduced in Section 2.1, where only a subset of grid nodes is observed at each day $t$, and the objective is to reconstruct the full temperature field $\boldsymbol{y}_t \in \mathbb{R}^N$ and quantify prediction confidence over all $N$ valid nodes.

Let $S_t \subset \{1, \dots, N\}$ denote the set of sensor locations available at day $t$, with $|S_t| = m$, and let $\boldsymbol{y}_{t,S_t}$ denote the corresponding observed temperatures. In the present work, the sensor set may be fixed (for a given placement design) or generated according to a placement strategy, but the reconstruction task remains the same: infer the temperature values at the unobserved nodes $\bar{S}_t$. The proposed framework maps sparse observations to two spatial outputs:

$$\left(\boldsymbol{y}_{t,S_t}, S_t\right) \mapsto (\widehat{\boldsymbol{\mu}}_t, \widehat{\boldsymbol{\sigma}}_t) \tag{4}$$

where $\widehat{\boldsymbol{\mu}}_t \in \mathbb{R}^N$ is the reconstructed mean temperature field and $\widehat{\boldsymbol{\sigma}}_t \in \mathbb{R}_+^N$ is the corresponding heteroscedastic predictive standard deviation field.

At a conceptual level, the framework consists of four coupled components. First, sparse observation encoding combines temperature values and a binary mask. The observed daily temperature values are embedded on the full grid by assigning measured values to sensor nodes and a neutral placeholder to unobserved nodes, while the mask explicitly distinguishes observed and unobserved nodes. This is essential in the temperature reconstruction setting because a value of zero in the encoded field should not be interpreted as a physical temperature measurement. Second, the valid grid nodes are represented as a spatial graph whose topology is constructed from geographic proximity (k-nearest neighbors in the spatial coordinates), optionally augmented by a global node to facilitate long-range information exchange. This graph formulation enables the model to exploit the spatial organization of the temperature field while remaining applicable to irregular valid-node masks. Third, a GAT-based reconstruction model uses the encoded sparse temperature observations, node coordinates (via positional encoding), and mask information as input, and predicts a full-field reconstruction in a normalized mean-residual form. The mean-residual parameterization stabilizes learning by conditioning the prediction on daily sensor statistics (sensor mean and standard deviation), which is particularly important for temperature fields with strong day-to-day shifts in level and spread. Fourth, in addition to the reconstructed mean temperature, a heteroscedastic uncertainty quantification head predicts a spatially varying uncertainty field through a log-variance output. This latent quantity is then transformed into predictive variance and predictive standard deviation for probabilistic evaluation and downstream decision-oriented products (e.g., exceedance probability maps).

Figure 1 summarizes this pipeline from sparse temperature measurements to reconstructed temperature and uncertainty maps. In practice, the framework is trained by maximizing the likelihood of unobserved temperatures conditioned on sparse observations, while enforcing physically reasonable spatial behavior through a mild regularization term (details in Sections 2.3 and 2.4). During inference, the same pipeline operates using only the sensor measurements and their locations for a given day $t$, and returns both $\hat{\boldsymbol{\mu}}_t$ and $\hat{\boldsymbol{\sigma}}_t$ over the entire temperature grid.

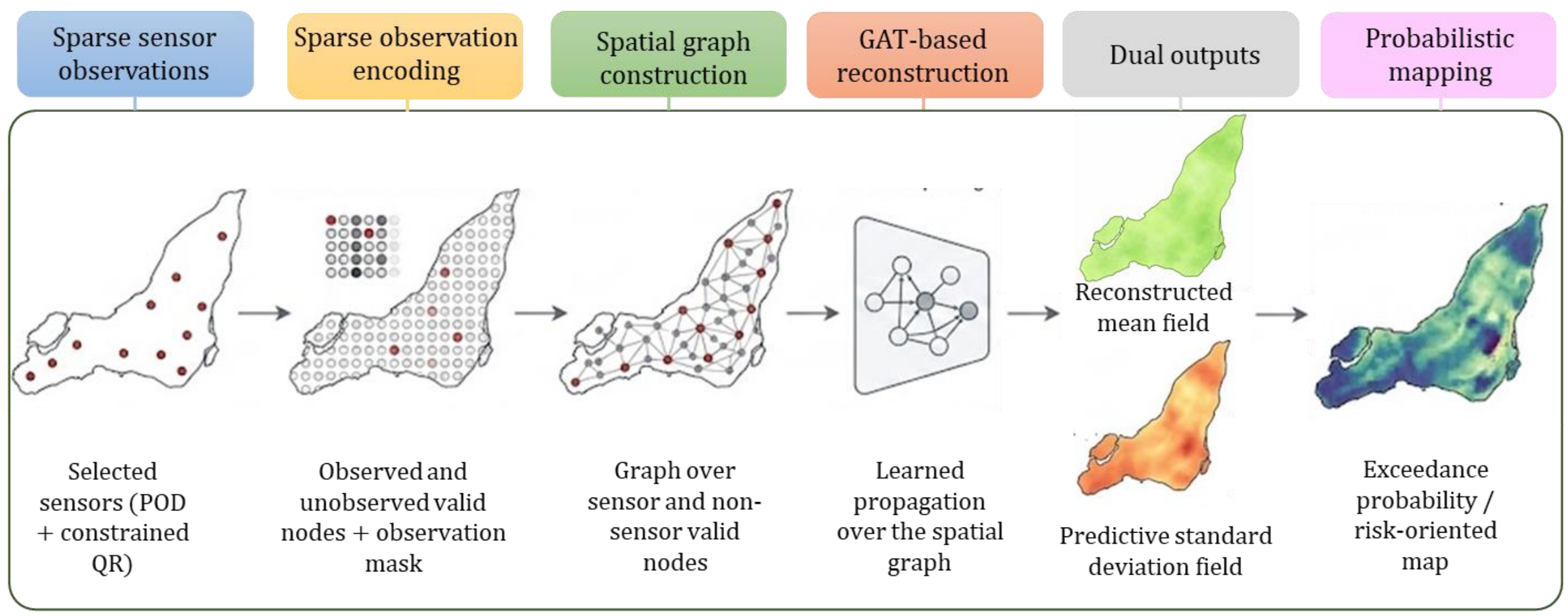

**Fig. 1** Method-centric pipeline for uncertainty-aware temperature field reconstruction from sparse sensors, including sparse observation encoding, spatial graph construction, GAT-based reconstruction, dual mean/uncertainty outputs, and downstream probabilistic mapping.

It is important to emphasize that the proposed framework is method-centric, in the sense that it specifies how sparse temperature observations are transformed into full-field predictions and uncertainty estimates. The application-specific evaluation protocol, including the temporal train/test split, comparisons with IDW and kriging baselines, and the use of uncertainty for decision-relevant exceedance mapping, is introduced later in the application and experimental sections. Similarly, sensor placement under a minimum-distance constraint is treated as a dedicated methodology component (Section 2.5) and is then evaluated in the application section through comparisons with alternative placement strategies.

## 2.3 Graph neural network for temperature field reconstruction

### 2.3.1 Graph construction for the temperature domain

Let $\{x_i\}_{i=1}^{N}$, with $x_i = (\lambda_i, \phi_i)$, denote the longitude-latitude coordinates of the $N$ valid grid nodes retained after masking invalid pixels. These nodes define the vertex set:

$$\mathcal{V} = \{1,2, \dots N\} \tag{5}$$

A spatial graph $\mathcal{G} = (\mathcal{V}, \mathcal{E})$ is constructed using a k-nearest-neighbor (kNN) rule in geographic coordinate space. For each node $i$, directed edges are created from its $K$ nearest neighbors (measured in the coordinate embedding used by the implementation), and the graph is then symmetrized to obtain an undirected adjacency. The resulting graph captures local spatial continuity in temperature fields while remaining compatible with irregular masks of valid nodes. To facilitate global information exchange (e.g., day-level temperature regime information that may not propagate efficiently through strictly local message passing), the graph is augmented with an additional global node $g$. The augmented vertex set is:

$$\tilde{\mathcal{V}} = \mathcal{V} \cup \{g\} \tag{6}$$

and bidirectional edges are added between the global node and every physical node (i.e., $(i, g) \in \tilde{\mathcal{E}}$, $(g, i) \in \tilde{\mathcal{E}}$, $\forall i \in \mathcal{V}$). The final graph used by the network is therefore $\tilde{\mathcal{G}} = (\tilde{\mathcal{V}}, \tilde{\mathcal{E}})$.

**2.3.2 Node features and sparse observation encoding**

For a given day $t$, let $S_t \subset \mathcal{V}$ denote the set of observed sensor nodes, with $|S_t| = m$. The observed temperatures are $\{y_{t,i}\}_{i \in S_t}$. To encode sparse observations on the full graph, we define a binary mask:

$$m_{t,i} = \begin{cases} 1, & i \in S_t \\ 0, & i \notin S_t \end{cases} \quad i \in \mathcal{V} \tag{7}$$

A direct embedding of sparse temperature measurements into a full-length vector is insufficient because unobserved nodes would be ambiguous (e.g., a placeholder value could be mistaken for a physical measurement). Therefore, the model uses both a value channel and a mask channel. To improve robustness across days with different mean temperature levels and amplitudes, the observed temperatures are normalized using sensor-wise day statistics $\bar{y}_{t,S_t} = \frac{1}{m}\sum_{i \in S_t} y_{t,i}$ and $s_{t,S_t} = \sqrt{\frac{1}{m-1}\sum_{i \in S_t}\left(y_{t,i} - \bar{y}_{t,S_t}\right)^2}$ with a lower bound applied to $s_{t,S_t}$ in implementation to prevent numerical instability on low-variance days. The normalized sparse input is then defined as:

$$z_{t,i} = \begin{cases} \frac{y_{t,i} - \bar{y}_{t,S_t}}{s_{t,S_t}}, & i \in S_t \\ 0, & i \notin S_t \end{cases} \tag{8}$$

In addition to the sparse temperature and mask channels, the model receives spatial positional information through a Fourier feature encoding of node coordinates. Let $\mathbf{p}_i \in \mathbb{R}^2$ denote the normalized coordinates of node $i$. A Fourier mapping $\gamma(\mathbf{p}_i) \in \mathbb{R}^{d_\gamma}$ is used to represent spatial position:

$$\gamma(\mathbf{p}_i) = [\sin(2\pi \mathbf{B}^{\mathrm{T}} \mathbf{p}_i), \cos(2\pi \mathbf{B}^{\mathrm{T}} \mathbf{p}_i)] \tag{9}$$

where $\mathbf{B}$ is a fixed random projection matrix (sampled once and kept fixed). A binary indicator $\delta_i$ is also included to distinguish physical nodes from the added global node $g$, defined as:

$$\delta_i = \begin{cases} 0, & i \in \mathcal{V} \\ 1, & i = g \end{cases} \tag{10}$$

The input feature vector for each node in the augmented graph is thus:

$$\mathbf{h}_{t,i}^{(0)} = [z_{t,i}, m_{t,i}, \gamma(\mathbf{p}_i), \delta_i], \quad i \in \tilde{\mathcal{V}} \tag{11}$$

with the convention that the global node receives $z_{t,g} = 0$ and $m_{t,g} = 0$. These features are linearly projected to the hidden dimension before graph message passing.

### 2.3.3 Mean-residual GAT reconstruction model

The reconstruction backbone is a multi-layer graph attention network (GAT)-type architecture with residual connections and normalization. Let $\mathbf{h}_{t,i}^{(\ell)} \in \mathbb{R}^{d_h}$ denotes the hidden representation of node $i$ at layer $\ell$. At each layer, node features are updated by aggregating messages from neighboring nodes in $\tilde{\mathcal{G}}$ using learned attention weights:

$$\mathbf{h}_{t,i}^{(\ell+1)} = \mathbf{h}_{t,i}^{(\ell)} + \mathcal{F}^{(\ell)}\left(\mathbf{h}_{t,i}^{(\ell)}, \left\{\mathbf{h}_{t,i}^{(\ell)} : j \in \mathcal{N}(i)\right\}\right) \tag{12}$$

where $\mathcal{N}(i)$ denotes the neighbors of node $i$, and $\mathcal{F}^{(\ell)}(.)$ represents the attention-based message passing block (including linear projections, multi-head attention aggregation, nonlinearity, and dropout). Layer normalization is applied within each block to stabilize training. The network predicts the normalized residual temperature field rather than the absolute temperature field directly. Specifically, the mean branch outputs $\hat{r}_{t,i} \in \mathbb{R}$ with $i \in \mathcal{V}$, which is interpreted as a residual in the normalized space defined by $\bar{y}_{t,S_t}$ and $s_{t,S_t}$. The reconstructed temperature means in physical units (℃) is obtained by de-normalization:

$$\hat{\mu}_{t,i} = \bar{y}_{t,S_t} + s_{t,S_t}\hat{r}_{t,i}, \quad i \in \mathcal{V} \tag{13}$$

This mean-residual parameterization is particularly important in the present temperature reconstruction setting because daily temperature fields can vary substantially in mean level across seasons and synoptic regimes, and centering by $\bar{y}_{t,S_t}$ removes much of this global variability from the learning task. In addition, scaling by $s_{t,S_t}$ normalizes the amplitude of spatial contrasts and gradients, making them easier to learn and reducing optimization sensitivity to warm versus cold days. Finally, the formulation ensures consistency with sparse observations by conditioning the reconstruction on observed sensor statistics, so that the model focuses on learning spatial structure rather than reconstructing absolute levels from scratch. The uncertainty branch is built on the same graph backbone and is introduced in Section 2.4, where the heteroscedastic predictive variance and the corresponding likelihood-based training formulation are defined.

### 2.4 Uncertainty quantification

In addition to reconstructing the mean temperature field, the proposed model provides a node-wise predictive uncertainty estimate for the daily temperature field conditioned on sparse observations. This is essential in the present problem because reconstruction quality is spatially nonuniform: uncertainty generally depends on the sensor configuration, the local spatial context, and the day-specific temperature pattern. Accordingly, we adopt a heteroscedastic formulation in which the predictive variance varies across nodes and days.

Let $\hat{\mu}_{t,i}$ denotes the reconstructed mean temperature at node $i \in \mathcal{V}$ on day $t$, as defined in Section 2.3.3. The uncertainty branch of the network predicts a scalar quantity for each node that is interpreted as the log-variance in the normalized residual space. Denoting this output by $\hat{\ell}_{t,i}^{(n)}$, the corresponding predictive variance in physical units (℃$^2$) is obtained through the same sensor-based scaling used in the mean-residual reconstruction:

$$\log \hat{\sigma}_{t,i}^2 = \hat{\ell}_{t,i}^{(n)} + 2 \log s_{t,S_t}, \quad i \in \mathcal{V} \tag{14}$$

where $s_{t,S_t}$ is the standard deviation of the observed sensor temperatures for day $t$ (Section 2.3.2). The predictive standard deviation is then:

$$\hat{\sigma}_{t,i} = \exp\left(\frac{1}{2} \log \hat{\sigma}_{t,i}^2\right) \tag{15}$$

This parameterization ensures positivity of the variance while preserving consistency with the mean-residual normalization. Given sparse observations $(\boldsymbol{y}_{t,S_t}, S_t)$, the temperature at each unobserved node $i \in \bar{S}_t$ is modeled as a Gaussian random variable:

$$y_{t,i} | \boldsymbol{y}_{t,S_t}, S_t \sim \mathcal{N}(\hat{\mu}_{t,i}, \hat{\sigma}_{t,i}^2) \tag{16}$$

This conditional Gaussian assumption yields a tractable likelihood and directly supports prediction intervals and exceedance probabilities in the application section. It also provides a principled mechanism for balancing reconstruction error against predictive sharpness: overconfident predictions are penalized through the likelihood when errors are large, while excessively diffuse predictions are penalized through the log-variance term.

Because the values at sensor nodes are given as inputs, training and evaluation should focus on the unobserved portion of the field. Let $\bar{S_t} = \mathcal{V} \backslash S_t$ denote the set of missing nodes on day $t$. The heteroscedastic Gaussian negative log-likelihood (NLL) for one day is written as:

$$\mathcal{L}_{\mathrm{NLL}}^{(t)} = \frac{1}{|\bar{S_t}|} \sum_{i \in \bar{S_t}} \frac{1}{2} \left[ \log(2\pi) + \log \hat{\sigma}_{t,i}^2 + \frac{\left(y_{t,i} - \hat{\mu}_{t,i}\right)^2}{\hat{\sigma}_{t,i}^2} \right] \tag{17}$$

The total data-fit term over a training set $\mathfrak{T}_{\mathrm{train}}$ is then:

$$\mathcal{L}_{\mathrm{NLL}} = \frac{1}{|\mathfrak{T}_{\mathrm{train}}|} \sum_{t \in \mathfrak{T}_{\mathrm{train}}} \mathcal{L}_{\mathrm{NLL}}^{(t)} \tag{18}$$

To ensure stable optimization, the predicted log-variance is bounded in implementation within a finite interval $\hat{\ell}_{t,i}^{(n)} \in [\ell_{\min}, \ell_{\max}]$ before de-normalization. This prevents pathological variance collapse ($\hat{\sigma}_{t,i} \longrightarrow 0$) or unrealistically large variances during training. A small positive floor may also be applied to $\hat{\sigma}_{t,i}$ during inference to avoid numerical singularities in downstream calculations (e.g., exceedance probabilities). In addition to the likelihood term, a mild spatial regularization is applied to the reconstructed mean field to discourage spurious high-frequency artifacts on the graph. Let $\mathcal{E}_{grid}$ denotes the set of edges among physical nodes (excluding the global node). A graph-Laplacian-type smoothness penalty is defined as:

$$\mathcal{L}_{\mathrm{smooth}}^{(t)} = \frac{1}{|\mathcal{E}_{grid}|} \sum_{(i,j) \in \mathcal{E}_{grid}} \left( \hat{\mu}_{t,i} - \hat{\mu}_{t,j} \right)^2 \tag{19}$$

The total training objective is:

$$\mathcal{L}_{\mathrm{train}} = \mathcal{L}_{\mathrm{NLL}} + \lambda_{\mathrm{smooth}} \mathcal{L}_{\mathrm{smooth}} \tag{20}$$

where $\lambda_{\mathrm{smooth}}$ is a small coefficient chosen so that the likelihood term remains dominant. In this work, the smoothness penalty is used only to regularize the mean reconstruction and does not directly constrain the uncertainty field.

The predicted standard deviation $\hat{\sigma}_{t,i}$ should be interpreted as a conditional predictive uncertainty associated with the reconstructed temperature at node $i$, given the sparse sensor observations and the trained model. In the present formulation, this uncertainty is learned through the heteroscedastic likelihood and is therefore primarily aimed at improving predictive calibration and supporting uncertainty-aware decision products.

Its practical quality is assessed in the application section using reliability diagnostics (coverage, CRPS, and NLL) on temporally held-out data.

### 2.5 Sensor placement under distance constraints

The reconstruction framework introduced in Sections 2.3-2.4 operates for any admissible set of sensor locations $S_t$. However, in practical deployments, sensor locations are not arbitrary: they are selected under operational, spatial, and budget constraints. In the present study, we focus on the minimum inter-sensor distance constraint, which prevents overly clustered deployments and promotes spatial coverage. Within this constrained setting, we adopt a POD-QR-based placement strategy to select informative sensor locations from the temperature field dynamics observed during the training period.

#### 2.5.1 Minimum-distance feasibility constraint

Let $d(i,j)$ denotes the geographic distance (in km) between nodes $i$ and $j$, computed from their latitude–longitude coordinates using a great-circle distance formula (Haversine distance in implementation). The placement is considered feasible if $d(i,j) \geq d_{min}, \forall i,j \in S, i \neq j$. This constraint serves two purposes. First, it reflects practical deployment considerations by avoiding redundant colocated sensors. Second, it imposes a physically meaningful lower bound on spacing, thereby facilitating fair comparisons among placement strategies at a fixed sensor budget $m$. If the requested $m$ is incompatible with the domain size and $d_{min}$, the feasible set may be reduced; in practice, this is handled by feasibility screening during the greedy selection process.

#### 2.5.2 POD basis construction

To select informative locations, we exploit the dominant spatial structure of the temperature field using Proper Orthogonal Decomposition (POD), computed on the training period to avoid temporal information leakage. Let $\mathfrak{T}_{\text{train}}$ denote the set of training days, and let $\boldsymbol{y}_t \in \mathbb{R}^N$ denote the full temperature field vector at day $t$. We assemble the temperature snapshot matrix $\mathbf{Y}_{\text{train}} = \left[\boldsymbol{y}_{t_1}^{\text{T}}, \boldsymbol{y}_{t_2}^{\text{T}}, \dots, \boldsymbol{y}_{t_{n_{tr}}}^{\text{T}}\right] \in \mathbb{R}^{n_{tr} \times N}$, with $t_k \in \mathfrak{T}_{\text{train}}$, where $n_{tr} = |\mathfrak{T}_{\text{train}}|$. A temporal mean is removed prior to decomposition so that the POD basis captures dominant spatial variability rather than only the climatological mean field $\widetilde{\mathbf{Y}}_{\text{train}} = \mathbf{Y}_{\text{train}} - \mathbf{1}\overline{\boldsymbol{y}}^T$ with $\overline{\boldsymbol{y}} \in \mathbb{R}^N$ the training-period mean across time and $\mathbf{1} \in \mathbb{R}^{n_{tr}}$ a vector of ones. An economy-size singular value decomposition is then performed:

$$\widetilde{\mathbf{Y}}_{\text{train}} = \mathbf{U}\mathbf{\Sigma}\mathbf{V}^{\text{T}} \tag{21}$$

The leading $r$ right singular vectors (columns of $\mathbf{V}$) define the retained POD basis $\mathbf{\Phi} = [\phi_1, \phi_2, \dots, \phi_r] \in \mathbb{R}^{N\times r}$, where $r$ is the retained POD rank. The rows of $\mathbf{\Phi}$ encode how strongly each candidate sensor location participates in the dominant temperature modes. This makes $\mathbf{\Phi}$ a natural matrix for selecting locations that are informative for reconstructing the principal spatial dynamics of the temperature field.

### 2.5.3 Constraint-aware POD–QR sensor selection

In the unconstrained setting, POD-based sensor placement is commonly performed by QR factorization with column pivoting applied to a matrix derived from $\mathbf{\Phi}$ (or equivalently by greedy maximization of linear independence among selected measurement rows). Here, we adapt this idea to enforce the minimum-distance feasibility constraint.

Let $\mathbf{\Phi}_{i,:} \in \mathbb{R}^r$ denote the $i$-th row of the POD basis matrix $\mathbf{\Phi}$, corresponding to node $i$. The goal is to select a set of rows (sensor locations) that are both: 1) informative (highly representative and linearly independent in the POD space), and 2) feasible under the spacing constraint. We adopt a greedy constrained selection procedure. At iteration $k$, let $S^{(k-1)}$ denote the set of already selected nodes. The candidate set is restricted to nodes that satisfy the distance constraint with respect to all previously selected sensors $\mathcal{C}^{(k)} = \{i \in \mathcal{V}\setminus S^{(k-1)}: d(i,j) \geq d_{min}, \forall j \in S^{(k-1)}\}$. Among feasible candidates, the next sensor is chosen to maximize a POD-space informativeness criterion consistent with pivoted-QR selection, i.e., preference is given to rows that contribute maximal new information relative to the span of previously selected rows. In implementation, this is achieved through a greedy residual-based row selection equivalent to a constrained pivoted-QR strategy. The process is repeated until $m$ sensors are selected or no feasible candidates remain. If no feasible candidate exists before reaching $m$, the algorithm terminates early and returns the feasible subset found so far. This behavior is explicitly reported when it occurs, since it reflects the interaction between domain geometry, sensor budget, and spacing constraint rather than a failure of the reconstruction model.

The placement strategy determines the sensor set $S$ used to generate sparse observations for field reconstruction. Once $S$ is selected, the graph neural reconstruction model and uncertainty quantification framework (Sections 2.3–2.4) operates exactly as defined, with no change in architecture or loss function. This separation is deliberate: it allows the reconstruction model and the placement strategy to be evaluated independently and jointly. In the application section, the proposed POD–QR placement is compared against alternative strategies (including random feasible placement and farthest-point sampling) under the same constraint and sensor budget. This enables a direct assessment of how sensor placement quality influences both reconstruction accuracy and uncertainty-aware temperature mapping.

# 3. Application and Results

## 3.1 Study domain, dataset, and experimental protocol

The proposed framework is evaluated for daily maximum near-surface air temperature field reconstruction over a Montréal-area polygonal domain using gridded meteorological data from Daymet Version 4 R1 (v4.1) (Thornton et al., 2022). Daymet provides daily surface weather estimates on a 1 km × 1 km grid in CF-compliant NetCDF format for North America and other supported regions, including daily maximum and minimum temperatures, precipitation, shortwave radiation, vapor pressure, snow water equivalent, and day length. In the present study, only the daily maximum temperature variable ($t_{max}$, in °C) is used. The selected Daymet product is particularly suitable for this application because it provides spatially continuous daily fields at a resolution compatible with urban- to regional-scale temperature mapping.

From the Daymet $t_{max}$ product, we extracted a Montreal polygon subset covering the period 1 January 2020 to 31 December 2024 at daily resolution. This yields 1825 daily snapshots (including leap years 2020 and 2024). The cropped domain is represented on a raw grid of 42 × 34 cells. Because the polygon mask excludes cells outside the study area, only valid cells are retained for reconstruction. To avoid temporal inconsistencies, a time-consistent valid mask is constructed by retaining only grid cells with finite values at all timesteps in the 2020–2024 record. After masking, the domain contains $N = 566$ valid nodes, and each daily temperature field is represented as a vector $\mathbf{y}_t \in \mathbb{R}^{566}$. This vectorized representation is used throughout the reconstruction, baseline comparison, and uncertainty analyses in Sections 3.2–3.5.

The experimental protocol follows a strict temporal hold-out design to assess generalization across time. Specifically, the data are split into a training period (2020–2023) and a test period (2024). The graph neural network (GNN) reconstruction model introduced in Section 2 is trained exclusively on the 2020–2023 subset, while all quantitative results reported in this section are computed on the 2024 hold-out set. To prevent information leakage in sensor placement, the POD basis used by the POD–QR placement strategy (Section 2.5) is also computed using the training period only. This separation is enforced consistently for reconstruction training, sensor placement construction, and uncertainty evaluation.

Sensor network design is evaluated under a minimum inter-sensor spacing constraint of 4 km (Haversine distance), reflecting a practical deployment requirement that discourages clustered measurements. Four sensor budgets are considered $m \in \{10,20,30,40\}$. For each budget, three placement strategies are compared: (i) random feasible placement (multiple seeds), (ii) farthest-point sampling (maximin coverage under the same spacing constraint), and (iii) constraint-aware POD–QR placement based on training-period temperature modes. Figure 2 illustrates representative layouts for these strategies at a fixed budget. Because the spacing constraint may limit the achievable number of sensors for some strategies and budgets, the actual selected count is tracked and reported during benchmarking.

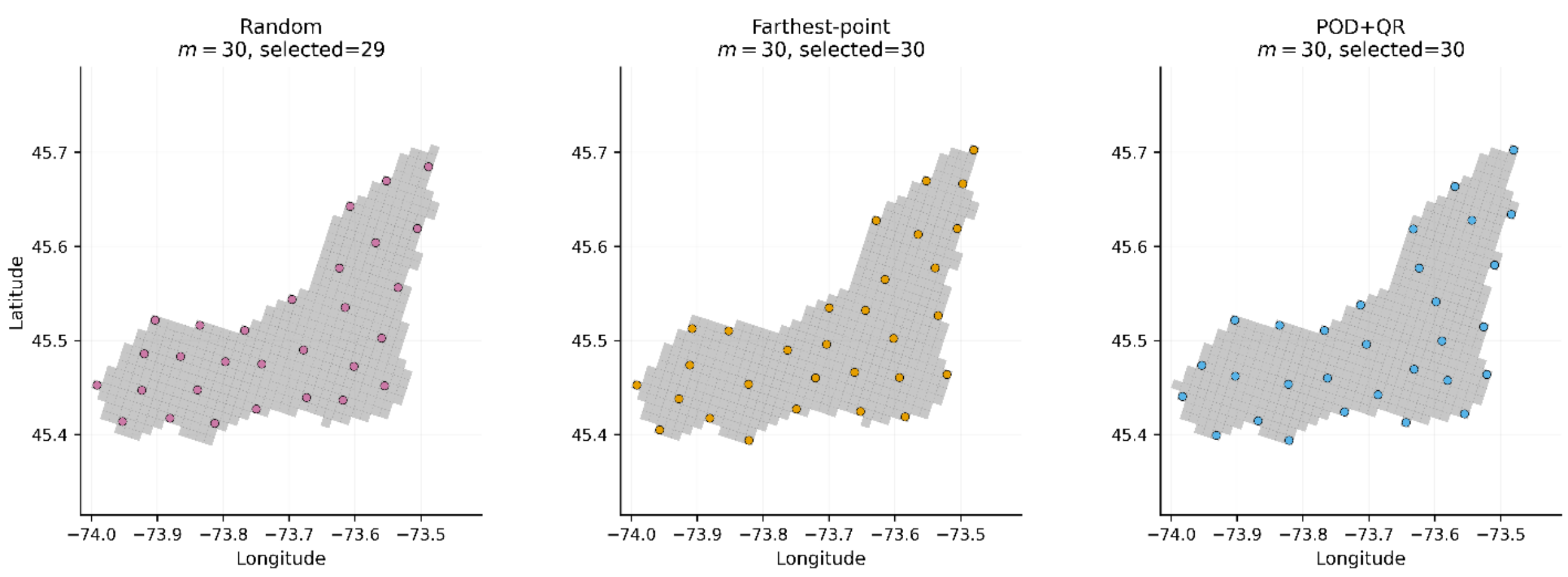

**Fig. 2** Representative sensor layouts under the 4 km minimum-spacing constraint

The reconstruction performance of the proposed uncertainty-aware GNN is compared against two classical interpolation baselines: inverse distance weighting (IDW) and ordinary kriging. In the implementation used for this study, IDW uses a fixed distance-power exponent, and kriging is implemented as ordinary kriging with an exponential variogram model. The comparison is intentionally performed under identical sensor sets so that differences in performance reflect the reconstruction method rather than the sensor placement. In Sections 3.2 and 3.4, results are reported both across sensor budgets and across placement strategies to isolate these two effects.

All quantitative metrics are computed on the unobserved (missing) nodes only, i.e., excluding sensor locations provided to the reconstruction model or interpolation baseline. The evaluation metrics include: (i) RMSE and MAE for deterministic reconstruction accuracy; (ii) negative log-likelihood (NLL) and continuous ranked probability score (CRPS) for probabilistic performance where uncertainty is available; (iii) empirical coverage at nominal 50%, 80%, 90%, and 95% prediction intervals together with reliability diagrams; and (iv) a Brier score for exceedance-probability evaluation in the decision-oriented mapping examples (Section 3.5). The corresponding comparative results are presented in the reconstruction accuracy plots and tables (Section 3.2), sensor-placement analyses (Section 3.3), uncertainty and calibration results (Section 3.4), and qualitative reconstruction/exceedance maps (Section 3.5).

### 3.2 Reconstruction accuracy

This subsection evaluates the deterministic reconstruction accuracy of the proposed uncertainty-aware GNN against two classical interpolation baselines, IDW and ordinary kriging, under a controlled sensor configuration. To isolate the effect of the reconstruction model (rather than the sensor layout), all three methods are first compared using the same sensor sets generated by the POD+QR placement strategy under the 4 km spacing constraint (Section 2.5). All metrics are computed on the unobserved nodes only in the 2024 hold-out period. Figure 3 shows a consistent ranking across all tested budgets ($m \in \{10,20,30,40\}$): the proposed GNN yields the lowest reconstruction error, followed by IDW, while kriging exhibits the largest error in the present configuration. This ordering is observed for both RMSE and MAE, indicating that the improvement is not limited to a specific error metric or sensitivity to outliers. The corresponding values summarized in Table 1 confirm the same trend quantitatively.

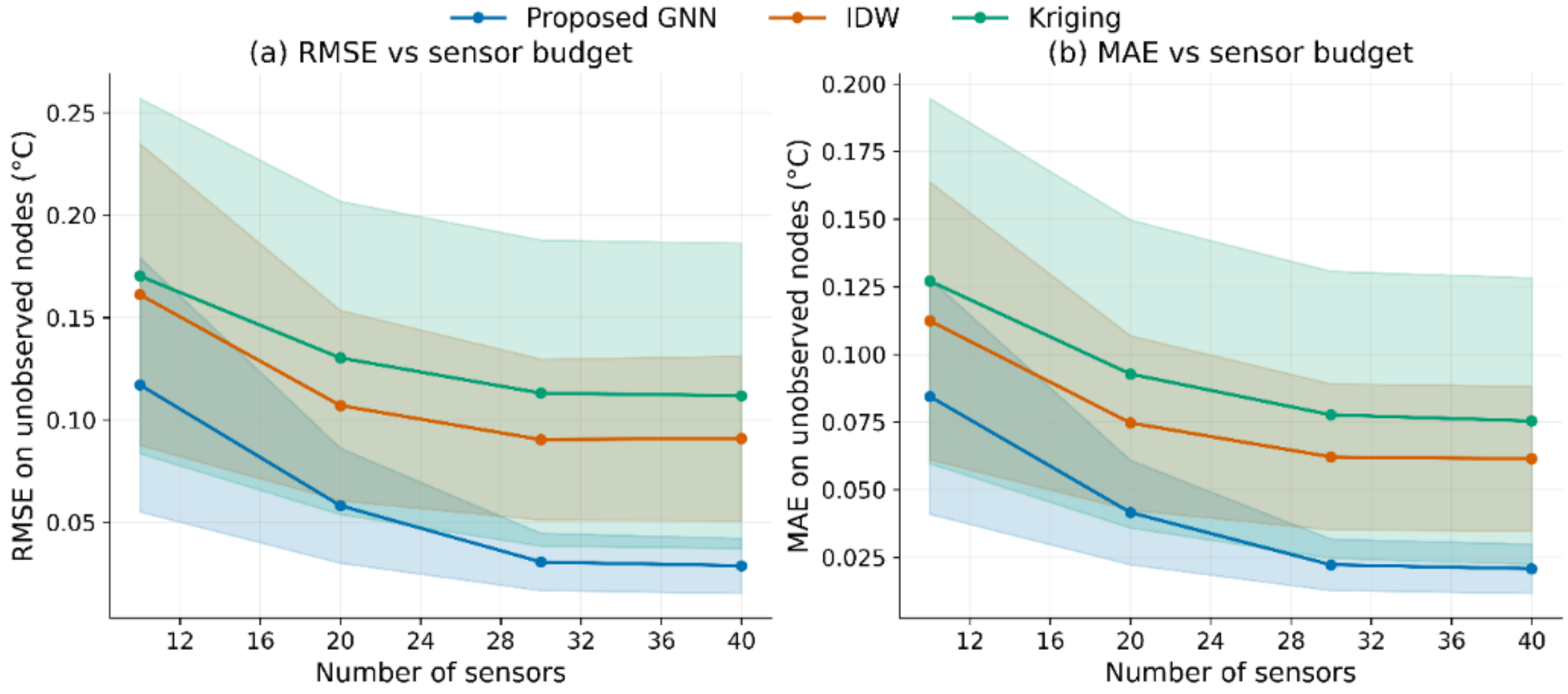

**Fig. 3** Reconstruction accuracy under POD+QR placement

A first notable result is the strong sensitivity of reconstruction error to sensor budget for the proposed GNN. Under POD+QR placement, the GNN RMSE decreases markedly as the number of sensors increases, with a large reduction between $m = 10$ and $m = 20$, followed by additional gains up to $m = 30$. The same behavior is visible in MAE. In contrast, both IDW and kriging improve more modestly with increasing sensor count and remain substantially less accurate than the GNN at all budgets. In the current results, the GNN reaches low-error reconstructions already around $m = 30$, whereas IDW and kriging remain in a significantly higher error regime, particularly for RMSE. This trend is consistent with the qualitative reconstruction maps presented later in Section 3.5, where the GNN better preserves the spatial temperature structure while limiting local reconstruction errors.

A second important observation is the emergence of diminishing returns beyond approximately 30 sensors. For the GNN, the gain from $m = 30$ to $m = 40$ is much smaller than the gain from $m = 10$ to $m = 20$, suggesting that the reconstruction problem is already well constrained near the $m \approx 30$ regime for the present domain and spacing constraint. This plateau effect is visible in both RMSE and MAE and provides an operationally relevant insight: increasing sensor density beyond this range may yield only marginal improvements in deterministic reconstruction accuracy. A similar flattening is also visible for the baseline methods, although at a much higher absolute error level. In the present results, this behavior is interpreted as a diminishing-returns regime, where once the domain becomes sufficiently constrained by the available observations, adding more sensors yields only marginal additional gains in reconstruction accuracy. Under the present domain geometry and 4 km spacing constraint, this practical saturation appears around $m \approx 30$.

The gap between the GNN and the classical interpolators becomes especially pronounced at moderate and high sensor budgets. This is a meaningful result from a methodological standpoint. IDW relies on a fixed distance-based weighting rule and cannot adapt to day-specific spatial patterns beyond geometric proximity. Ordinary kriging, while statistically principled, remains tied to a prescribed covariance/variogram structure and may be limited in the presence of non-stationarity, trend effects, or spatial behaviors that are not well captured by the selected variogram model. By contrast, the proposed GNN learns a nonlinear mapping from sparse observations to the full field while explicitly exploiting graph connectivity and positional encoding, which appears to translate into superior temperature reconstruction on the temporally held-out year.

The standard deviations reported in Table 1 and illustrated in figure 3 indicate that variability across test days is non-negligible, as expected for daily temperature fields spanning different seasons and synoptic conditions. Nevertheless, the separation between the GNN and the two baselines remains robust relative to this day-to-day variability. In other words, the observed ranking is not driven by a small subset of favorable days, but is persistent across the tested 2024 conditions. Overall, the results in Fig. 3 and Table 1 demonstrate that the proposed graph-based reconstruction framework provides a clear deterministic accuracy advantage over IDW and ordinary kriging across all tested sensor budgets under the same placement strategy. This establishes the proposed GNN as the strongest reconstruction method among the compared approaches before examining, in Section 3.3, how the choice of sensor placement strategy further affects performance under the 4 km spacing constraint.

**Table 1** RMSE and MAE on unobserved nodes for the 2024 test set under the 4 km spacing constraint (mean ± standard deviation across days), comparing the proposed GNN, IDW, and kriging across sensor budgets and placement strategies.

| placement | budget | GNN RMSE | GNN MAE | IDW RMSE | IDW MAE | Kriging RMSE | Kriging MAE |
|---|---|---|---|---|---|---|---|
| POD+QR | 10 | 0.117 ± 0.062 | 0.084 ± 0.044 | 0.161 ± 0.074 | 0.112 ± 0.051 | 0.170 ± 0.087 | 0.127 ± 0.067 |
| | 20 | 0.058 ± 0.028 | 0.042 ± 0.019 | 0.107 ± 0.046 | 0.075 ± 0.032 | 0.130 ± 0.076 | 0.093 ± 0.057 |
| | 30 | 0.031 ± 0.014 | 0.022 ± 0.009 | 0.090 ± 0.039 | 0.062 ± 0.027 | 0.113 ± 0.075 | 0.078 ± 0.053 |
| | 40 | 0.029 ± 0.013 | 0.021 ± 0.009 | 0.091 ± 0.040 | 0.061 ± 0.027 | 0.112 ± 0.075 | 0.075 ± 0.053 |
| Random | 10 | 0.102 ± 0.060 | 0.068 ± 0.039 | 0.152 ± 0.068 | 0.099 ± 0.045 | 0.150 ± 0.070 | 0.099 ± 0.047 |
| | 20 | 0.047 ± 0.025 | 0.031 ± 0.015 | 0.130 ± 0.058 | 0.081 ± 0.035 | 0.120 ± 0.062 | 0.072 ± 0.039 |
| | 30 | 0.032 ± 0.017 | 0.021 ± 0.010 | 0.108 ± 0.051 | 0.068 ± 0.030 | 0.107 ± 0.065 | 0.065 ± 0.043 |
| | 40 | 0.031 ± 0.017 | 0.021 ± 0.010 | 0.108 ± 0.051 | 0.067 ± 0.030 | 0.106 ± 0.065 | 0.065 ± 0.043 |
| Farthest-point | 10 | 0.128 ± 0.061 | 0.085 ± 0.039 | 0.168 ± 0.082 | 0.110 ± 0.051 | 0.166 ± 0.086 | 0.108 ± 0.055 |
| | 20 | 0.038 ± 0.016 | 0.028 ± 0.012 | 0.102 ± 0.047 | 0.073 ± 0.034 | 0.138 ± 0.085 | 0.098 ± 0.062 |
| | 30 | 0.030 ± 0.014 | 0.021 ± 0.009 | 0.080 ± 0.034 | 0.058 ± 0.025 | 0.113 ± 0.080 | 0.078 ± 0.058 |
| | 40 | 0.030 ± 0.014 | 0.021 ± 0.009 | 0.080 ± 0.034 | 0.058 ± 0.025 | 0.113 ± 0.080 | 0.078 ± 0.058 |

### 3.3 Effect of sensor placement under spacing constraint

This subsection analyzes the influence of sensor placement strategy on reconstruction performance under the minimum inter-sensor spacing constraint of 4 km. To isolate the effect of placement rather than reconstruction methodology, the comparison is performed using the proposed GNN reconstruction model while varying sensor locations according to the three strategies defined in Section 2.5: random feasible placement, farthest-point sampling, and constraint-aware POD+QR placement. All results are computed over the 2024 hold-out period and evaluated exclusively on unobserved nodes.

Figure 4 reports RMSE and MAE versus sensor budget for the three placement strategies. The results show that sensor placement has a clear impact on reconstruction quality, especially at low sensor budgets, but that the magnitude and direction of this effect depend on the budget regime. In particular, the differences among strategies are largest at $m = 10$, remain noticeable at $m = 20$, and become much smaller at $m = 30$ and $m = 40$, where the performance curves converge.

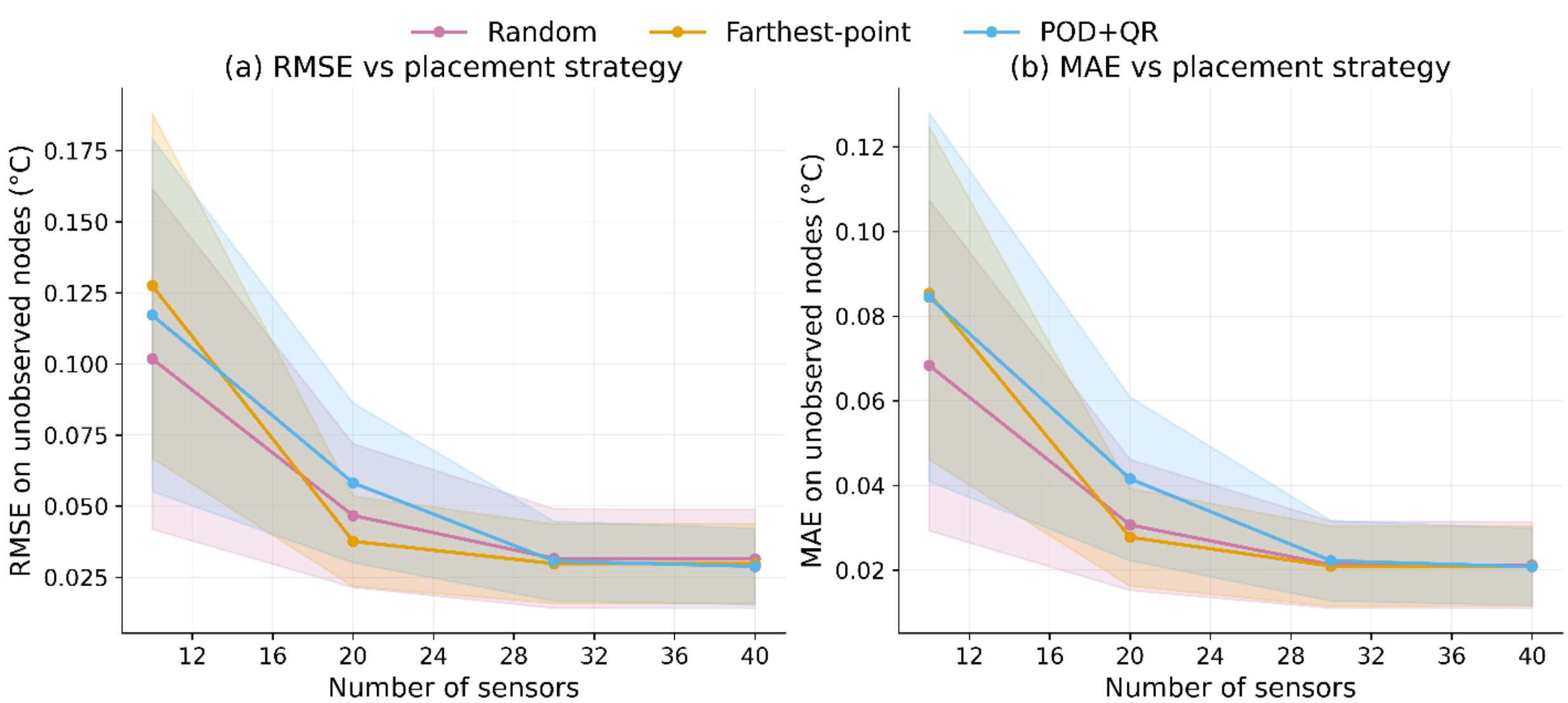

**Fig. 4** Effect of sensor placement strategy under the 4 km spacing constraint (Proposed GNN)

At the lowest budget ($m = 10$), the random feasible placement yields the lowest average reconstruction error in the present experiments, outperforming both POD+QR and farthest-point sampling. This is an important outcome because it indicates that, under severe sparsity and a distance-constrained domain, maximizing modal informativeness (POD+QR) or maximizing geometric separation (farthest-point) does not automatically guarantee the best reconstruction for the learned GNN. One plausible explanation is that the GNN is trained with randomly sampled observation masks, which reduces reliance on any single fixed sensor configurations and improves robustness to varying sparse sensor layouts. This may favor generalization to random feasible placements, especially when the number of sensors is too small for any strategy to resolve all dominant spatial modes reliably. In addition, random feasible placement may occasionally produce favorable sensor allocations with better alignment to day-specific temperature gradients than a single deterministic design.

At the intermediate budget ($m = 20$), farthest-point sampling becomes highly competitive and in the current results achieves the best average error. This suggests that, once a minimum number of sensors is available, enforcing broad geometric coverage can provide a strong advantage by reducing large unsensed regions and improving the conditioning of the reconstruction problem. In contrast, POD+QR and random placement remain close but slightly inferior on average for this budget. This behavior highlights the trade-off between coverage-oriented design (farthest-point) and mode-informativeness-oriented design (POD+QR), and indicates that the optimal design criterion may depend on the available sensor budget.

For $m = 30$ and $m = 40$, the three placement strategies yield very similar RMSE and MAE values, with only small differences relative to the low-budget regime. This convergence indicates that the reconstruction problem becomes sufficiently constrained as sensor density increases, so that the placement strategy exerts a weaker influence on the final error. From an operational perspective, this is a useful finding: under the present domain geometry and 4 km spacing constraint, the benefit of optimizing placement decreases once the sensor network reaches a moderate density. Combined with the plateau observed in Section 3.2, this suggests a practical “saturation regime” in which additional design effort or additional sensors may produce only marginal gains.

The representative sensor layouts shown in Fig. 2 (for a fixed budget, e.g., $m = 30$) further clarify the geometric behavior of the compared strategies. As expected, farthest-point sampling emphasizes spatial spread, POD+QR emphasizes modal informativeness under feasibility constraints, and random feasible placement produces a stochastic but admissible distribution. The corresponding performance trends in Fig. 4 confirm that these geometric differences matter most when the sensor budget is small, whereas at higher budgets the domain is sufficiently sampled for the GNN to reconstruct the temperature field accurately across all three placement strategies. It is also important to interpret these results in the context of the 4 km minimum-distance constraint. This constraint limits the feasible sensor configurations and can reduce the effective difference between strategies as the budget increases. In some cases, the feasible solution set may become sufficiently restricted that multiple strategies produce sensor networks with similar coverage characteristics, which contributes to the observed performance convergence. For transparency, the actual number of selected sensors is tracked in the benchmarking pipeline, enabling verification that the requested sensor budget is feasible under each strategy and helping explain any apparent performance plateaus.

### 3.4 Uncertainty quantification performance and calibration

This subsection evaluates the probabilistic performance of the proposed heteroscedastic GNN and compares it with the uncertainty estimates provided by ordinary kriging. Kriging is used as the baseline in this subsection because, unlike standard IDW, it provides a predictive uncertainty estimate and therefore enables direct comparison of calibration-oriented metrics such as coverage, reliability, CRPS, and NLL. The analysis focuses on the quality of predictive uncertainty in the 2024 hold-out period, using the metrics introduced in Section 3.1: negative log-likelihood (NLL), continuous ranked probability score (CRPS), and empirical coverage at nominal 50%, 80%, 90%, and 95% prediction intervals. The uncertainty comparisons shown in the subsequent figures and table are reported under the POD+QR placement strategy so that differences reflect uncertainty modeling performance rather than sensor geometry.

#### 3.4.1 Calibration behavior: reliability and empirical coverage

Figure 5 compares the reliability behavior of the GNN and kriging through nominal-versus-empirical coverage curves for multiple sensor budgets. A clear contrast emerges between the two methods. Points on the diagonal indicate perfect calibration; curves below the diagonal indicate under-coverage (overconfident intervals), whereas curves above the diagonal indicate over-coverage (overly wide intervals). For the GNN, the empirical coverage curves move progressively toward the ideal diagonal as the sensor budget increases from $m = 10$ to $m = 40$. At low budget ($m = 10$), the GNN exhibits under-coverage, particularly at the lower nominal levels, indicating that the predictive intervals are somewhat too narrow in the strongly sparse regime. This is consistent with the increased difficulty of reconstructing the field from very limited observations. As the number of sensors increases ($m = 20,30,40$), the coverage curves shift upward toward the target levels and become substantially better aligned with the diagonal. The corresponding coverage values in Table 2 confirm this trend: the GNN achieves near-target coverage at the upper interval levels (especially 90% and 95%) for the larger sensor budgets, while maintaining significantly improved calibration relative to the $m = 10$ case.

In contrast, the kriging reliability curves lie systematically above the diagonal across the tested budgets, indicating persistent over-coverage (i.e., prediction intervals that are too wide). This over-conservative behavior is also evident in the empirical coverage values in Table 2, where kriging frequently exceeds the nominal target by a substantial margin, especially at the lower nominal levels (e.g., nominal 50% intervals yielding empirical coverage far above 0.50). From a practical standpoint, this implies that kriging uncertainty in the present setup is insufficiently sharp, which can reduce the discriminative value of probabilistic outputs for downstream decision-making.

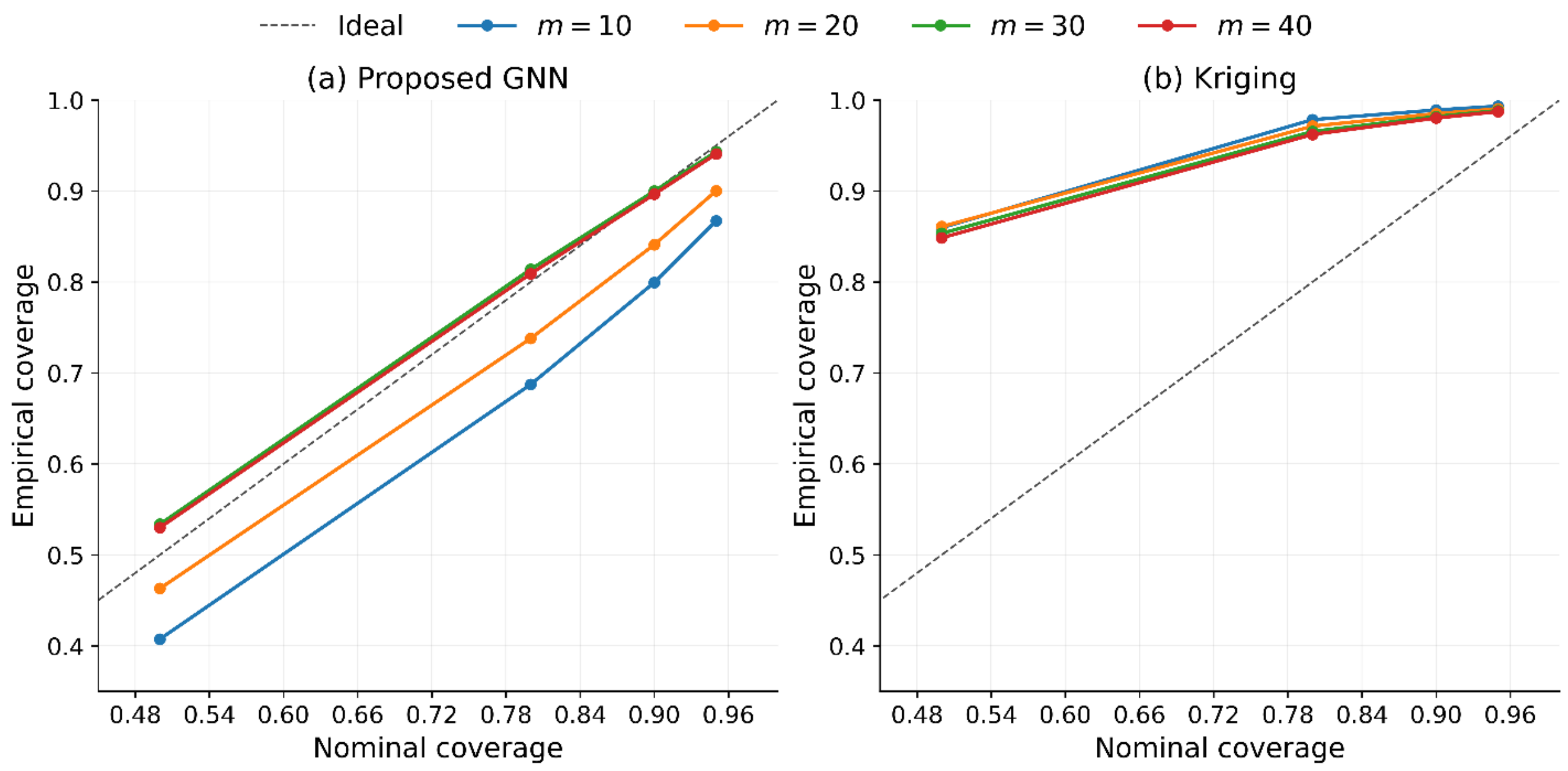

**Fig. 5** Reliability diagrams under POD+QR placement

### 3.4.2 Sharpness–calibration trade-off: CRPS and NLL

Figure 6 reports CRPS and NLL as functions of sensor budget for the GNN and kriging. These metrics jointly assess probabilistic performance by penalizing both inaccurate means and mis-scaled uncertainties. Lower CRPS indicates better probabilistic forecasts, while lower NLL (more negative values in the present setting) indicates higher likelihood assigned to the observed temperatures at unobserved nodes.

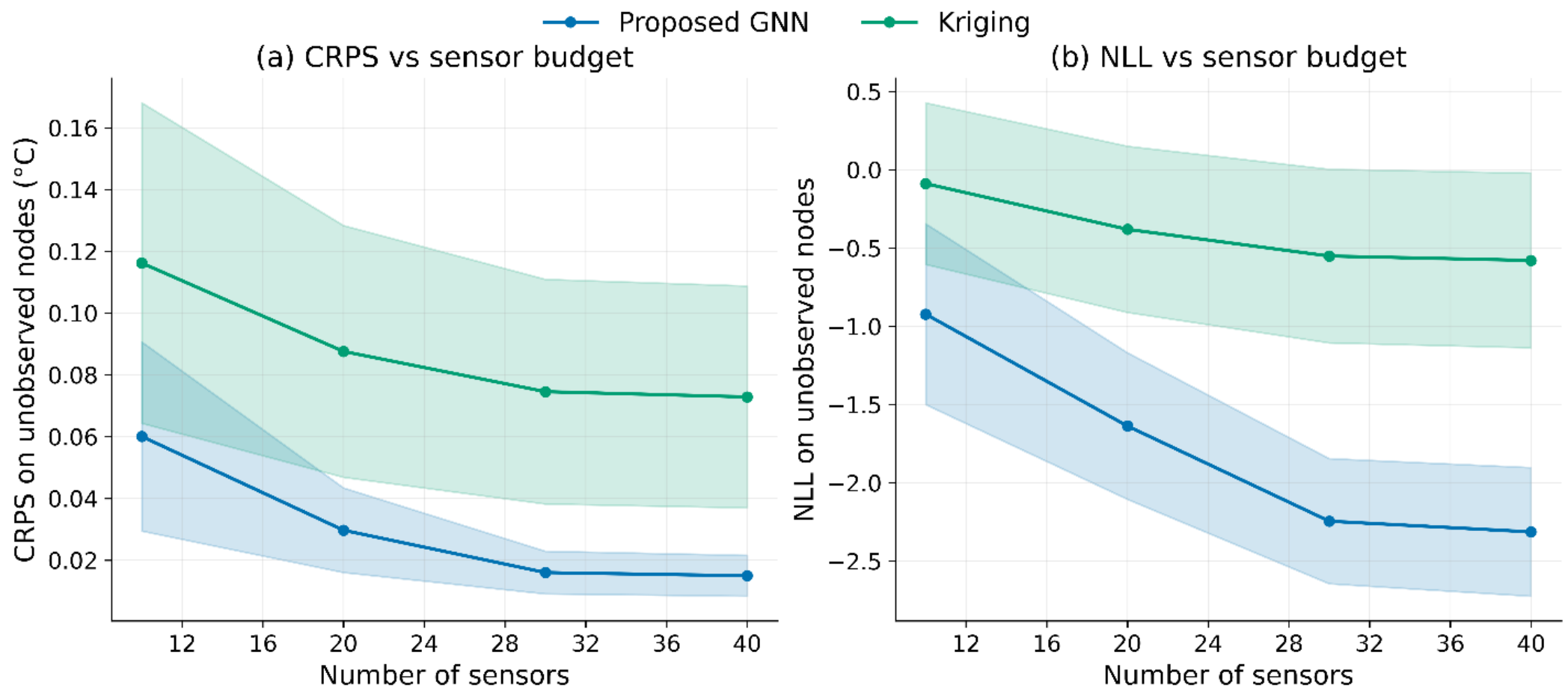

**Fig. 6** Probabilistic performance under POD+QR placement

The proposed GNN consistently achieves lower CRPS than kriging across all tested sensor budgets. Moreover, GNN CRPS decreases steadily as the number of sensors increases, reflecting both improved mean reconstruction and more informative uncertainty estimates. Kriging CRPS also improves with additional sensors, but remains substantially higher than the GNN values throughout. This gap indicates that the proposed model not only reconstructs the mean more accurately as already reflected by the lower RMSE and MAE values in Section 3.2 (Fig. 3 and Table 1), but also translates that deterministic advantage into superior probabilistic predictions. A similar pattern is observed for NLL. The GNN yields markedly lower NLL values than kriging across the tested budgets, with performance improving as sensor density increases. Kriging NLL decreases more slowly and remains substantially higher than the GNN values, consistent with the over-coverage observed in the reliability analysis. In probabilistic terms, this suggests that the kriging predictive distributions are too diffuse to achieve competitive likelihood scores, even when their empirical coverage appears "safe" or conservative. By contrast, the GNN achieves a better balance between calibration and sharpness, producing predictive intervals that are both more informative and better aligned with observed error behavior.

These trends are reinforced by the uncertainty summary statistics in Table 2, which show coherent improvements in GNN coverage, CRPS, and NLL from $m = 10$ to $m = 40$, and systematically weaker probabilistic performance for kriging in the current implementation.

### 3.4.3 Implications for uncertainty-aware temperature mapping

The uncertainty results in Figs. 5–6 and Table 2 support two key conclusions. First, the proposed heteroscedastic GNN provides meaningful and progressively better-calibrated uncertainty estimates as sensor density increases, rather than merely producing deterministic reconstructions with heuristic confidence values. Second, the probabilistic advantage of the GNN over kriging is not limited to a single metric: it is observed simultaneously in reliability behavior (coverage) and proper scoring rules (CRPS and NLL).

This is particularly important for the decision-oriented analyses considered later in Section 3.5, where uncertainty is used to construct exceedance probability maps. In that context, over-dispersed uncertainty (as observed for kriging) can lead to overly smooth or weakly discriminative probability fields, while under-calibrated uncertainty can yield overconfident predictions. The results presented here indicate that the proposed GNN offers a more favorable uncertainty representation for such applications, especially at moderate to high sensor budgets where both reconstruction accuracy and calibration improve substantially.

**Table 2** Uncertainty performance on unobserved nodes for the 2024 test set under POD+QR placement and the 4 km spacing constraint (mean ± standard deviation across days), comparing the proposed GNN and kriging.

| Budget | GNN | | | | | | | Kriging | | | | | | |
|---|---|---|---|---|---|---|---|---|---|---|---|---|---|---|
| | **CRPS** | **NLL** | **Cov50** | **Cov80** | **Cov90** | **Cov95** | **Brier** | **CRPS** | **NLL** | **Cov50** | **Cov80** | **Cov90** | **Cov95** | **Brier** |
| 10 | 0.060 ± 0.031 | -0.922 ± 0.578 | 0.407 ± 0.116 | 0.687 ± 0.150 | 0.799 ± 0.136 | 0.867 ± 0.115 | 0.00093 ± 0.01073 | 0.116 ± 0.052 | -0.088 ± 0.516 | 0.860 ± 0.106 | 0.978 ± 0.028 | 0.989 ± 0.012 | 0.993 ± 0.007 | 0.00158 ± 0.01413 |
| 20 | 0.030 ± 0.014 | -1.637 ± 0.468 | 0.463 ± 0.105 | 0.738 ± 0.120 | 0.841 ± 0.104 | 0.900 ± 0.085 | 0.00032 ± 0.00359 | 0.088 ± 0.041 | -0.380 ± 0.531 | 0.861 ± 0.101 | 0.971 ± 0.025 | 0.985 ± 0.014 | 0.990 ± 0.009 | 0.00108 ± 0.01047 |
| 30 | 0.016 ± 0.007 | -2.245 ± 0.400 | 0.534 ± 0.096 | 0.814 ± 0.098 | 0.900 ± 0.079 | 0.943 ± 0.060 | 0.00014 ± 0.00140 | 0.075 ± 0.036 | -0.551 ± 0.555 | 0.853 ± 0.117 | 0.965 ± 0.029 | 0.982 ± 0.016 | 0.988 ± 0.010 | 0.00083 ± 0.00831 |
| 40 | 0.015 ± 0.007 | -2.313 ± 0.411 | 0.530 ± 0.095 | 0.809 ± 0.099 | 0.896 ± 0.081 | 0.941 ± 0.063 | 0.00014 ± 0.00145 | 0.073 ± 0.036 | -0.579 ± 0.558 | 0.849 ± 0.122 | 0.962 ± 0.032 | 0.980 ± 0.017 | 0.987 ± 0.012 | 0.00077 ± 0.00792 |

## 3.5 Qualitative field reconstruction and exceedance probability mapping

This subsection complements the aggregate metrics reported in Sections 3.2–3.4 by examining representative heat-event reconstructions and the corresponding probabilistic exceedance maps. The goal is to assess whether the proposed framework produces spatially coherent mean and uncertainty fields on individual high-temperature days, and whether these outputs translate into interpretable decision-oriented products.

Figure 7 presents qualitative temperature-field reconstruction results for the 2024-06-18 and 2024-06-19 heat event using a fixed POD+QR sensor configuration with $m = 30$ sensors. For each date, four maps are shown: the ground-truth field, the GNN reconstructed mean field, the absolute error field, and the predicted standard deviation field. The reconstructed mean fields reproduce the dominant spatial temperature gradients and localized structures visible in the ground truth, while the absolute error remains low over most of the domain. The error maps indicate that residual discrepancies are spatially concentrated rather than systematic, which is consistent with the low RMSE and MAE values reported for the $m = 30$ budget in Sections 3.2–3.3. The predicted standard deviation fields are spatially nonuniform and exhibit higher values in localized regions, indicating that the model uncertainty responds to spatial reconstruction difficulty rather than acting as a spatially constant confidence band. This behavior is consistent with the calibration and CRPS/NLL trends discussed in Section 3.4.

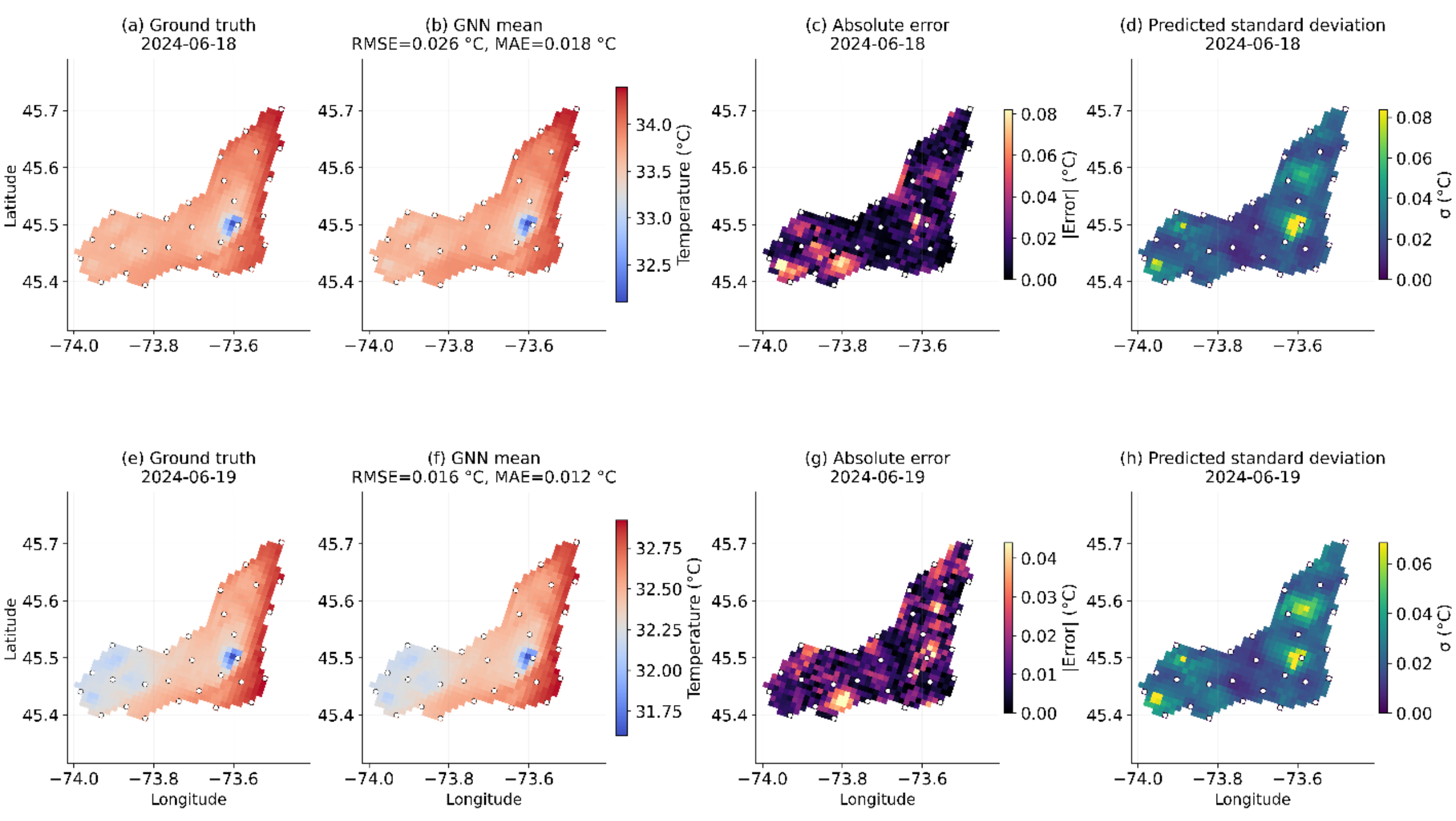

**Fig. 7** Qualitative temperature field reconstruction for the 2024-06-18/2024-06-219 heat event (POD+QR, m=30)

Figure 8 illustrates the decision-oriented use of the predicted uncertainty through exceedance probability mapping for 2024-06-19 at an application-relevant threshold of 32.5 °C. The figure includes (a) the ground-truth exceedance indicator (binary field), (b) the GNN exceedance probability map, and (c) the kriging exceedance probability map. It is important to note that the ground-truth exceedance map is binary by definition (0 or 1), whereas the GNN and kriging maps are probabilistic estimates. Accordingly, intermediate probabilities are expected in regions near the exceedance boundary or in areas of higher predictive uncertainty. The objective is therefore not visual equality with the binary ground truth, but rather whether the probability field is spatially consistent with the exceedance pattern and whether it is sufficiently discriminative.

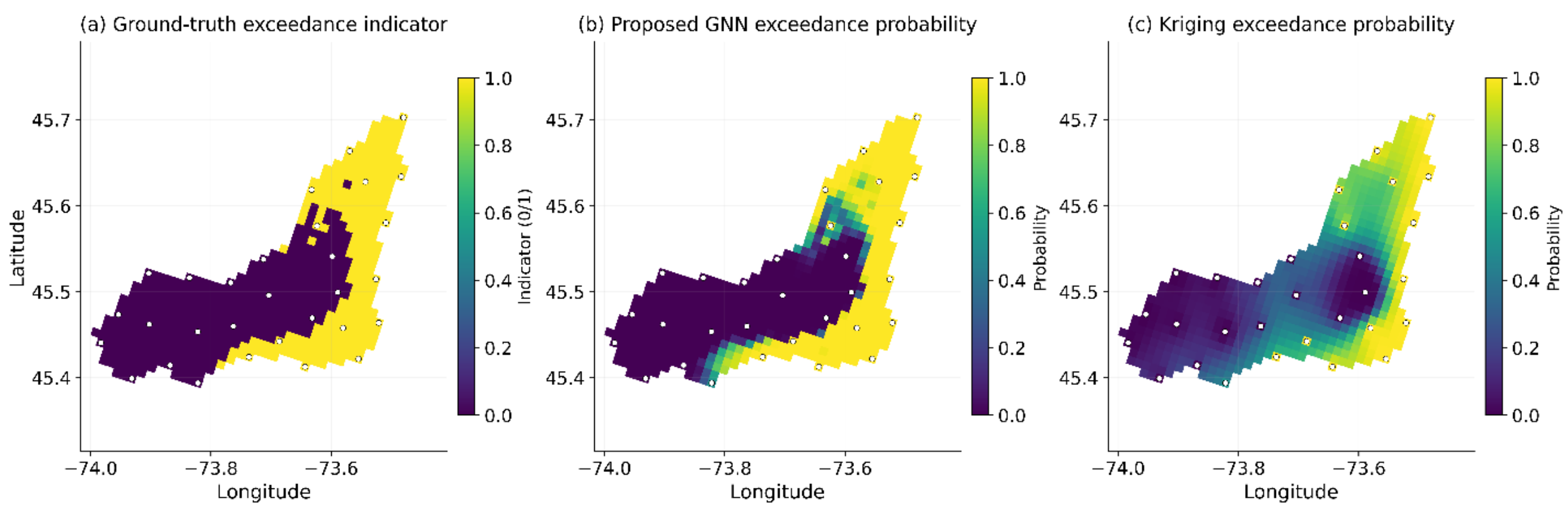

**Fig. 8** Exceedance probability mapping for 2024-06-19 (threshold = 32.5 °C)

In the present case, the GNN exceedance probability map exhibits a spatially coherent and relatively sharp transition between low- and high-probability regions, with intermediate probabilities concentrated near the apparent exceedance boundary and in localized uncertain zones. This is consistent with the structured uncertainty fields observed in Fig. 7 and with the improved probabilistic performance (CRPS/NLL and calibration behavior) reported in Section 3.4. By contrast, the kriging exceedance probability map displays a broader spread of intermediate probabilities across a larger portion of the domain, yielding a more diffuse transition pattern. This behavior is consistent with the over-conservative coverage tendencies observed for kriging in Fig. 7 and Table 2, and reflects a less discriminative uncertainty representation in the current setup.

## 4. Discussion

The results presented in Section 3 show that the proposed uncertainty-aware graph neural network (GNN) provides a consistent advantage for daily temperature field reconstruction under sparse sensing, while also yielding practically useful predictive uncertainty. Beyond the metric-level comparisons, several methodological and application-oriented implications emerge from these results.

A first major outcome is the robust superiority of the GNN over IDW and ordinary kriging for deterministic reconstruction accuracy across the tested sensor budgets. This improvement is observed in both RMSE and MAE and persists across the 2024 hold-out period, indicating that the model has learned a transferable mapping rather than overfitting to day-specific patterns in the training set. The performance gap is particularly pronounced at moderate to high sensor budgets, where the GNN appears able to exploit the richer observation patterns more effectively than the fixed-form interpolation baselines. In the present setting, this likely reflects the ability of the graph-based model to represent nonlinear spatial relationships and to integrate both local and global context (through graph connectivity, positional encoding, and the global node), whereas IDW and kriging remain constrained by geometric weighting rules or prescribed covariance structures.

At the same time, the results also suggest that baseline performance is strongly implementation- and assumption-dependent, especially for kriging. In the current experiments, kriging underperforms both the GNN and IDW in deterministic accuracy and exhibits over-conservative uncertainty behavior. While this does not diminish the observed advantage of the proposed method, it should be interpreted in light of the chosen kriging configuration (ordinary kriging with a specified variogram model and the coordinate handling adopted in the implementation). Temperature fields over urban domains may exhibit nonstationarity, anisotropy, or trend components that are not fully captured by a single stationary variogram model. Accordingly, the present comparison demonstrates strong practical gains over a standard kriging setup, while also motivating future sensitivity analyses (e.g., variogram families, anisotropy treatment, detrending) if one wishes to further optimize the kriging baseline.

A second important finding concerns the role of sensor placement under the 4 km minimum-distance constraint. The results show that placement matters most in the sparse regime, but that the ranking among placement strategies is not uniform across budgets. In particular, random feasible placement and farthest-point sampling can outperform constraint-aware POD+QR placement at low or intermediate budgets in the present experiments, while differences shrink substantially at higher budgets. This is a valuable result because it underscores that sensor-placement quality is not absolute: it depends on the deployment budget, the spatial constraint set, and the downstream reconstruction model. In other words, a placement strategy that is optimal under one criterion (e.g., modal observability in POD space) may not remain optimal once combined with a nonlinear learned reconstruction model trained on random observation patterns. The observed budget dependence therefore supports evaluating placement strategies jointly with the intended inference model rather than treating sensor design as a method-agnostic preprocessing step.

The results also reveal a clear diminishing-returns regime around $m \approx 30$ sensors for the present domain and spacing constraint. This plateau is visible in deterministic accuracy (RMSE/MAE) and in probabilistic scores (CRPS/NLL), and is reinforced by the qualitative reconstruction maps in Section 3.5. From an operational perspective, this is one of the most useful outcomes of the study: under the tested configuration, increasing the network size beyond this range yields only marginal improvements. Such information is directly relevant for deployment planning, where sensor procurement, installation, maintenance, and telemetry costs must be balanced against expected gains in reconstruction quality. In this sense, the framework provides not only a reconstruction tool but also a means to identify practical sensor-budget regimes with favorable cost–benefit characteristics.

The uncertainty analysis is another central contribution of the study. The proposed GNN does not merely produce point estimates; it provides heteroscedastic, spatially varying uncertainty estimates that improve in calibration as sensor density increases. The reliability diagrams show that the GNN is under-confidently calibrated at low budget only to a limited extent (under-coverage at small nominal intervals), but progressively approaches the ideal calibration line for larger budgets. This behavior is consistent with the reconstruction task becoming better constrained as more observations are provided. By contrast, kriging exhibits persistent over-coverage in the current setup, indicating over-dispersed predictive distributions. The CRPS and NLL results reinforce this interpretation by showing that the GNN attains a better sharpness–calibration trade-off than kriging across all tested budgets. This is particularly significant because proper scoring rules penalize both overconfident and overly diffuse predictions; thus, the GNN's advantage reflects genuine probabilistic quality rather than simply broader intervals.

The practical value of this uncertainty modeling is illustrated in the exceedance probability mapping examples. The key lesson from these examples is methodological as much as visual: meaningful comparison of probabilistic exceedance maps requires informative threshold/day combinations. All-zero or all-one exceedance cases are mathematically valid but visually degenerate and can obscure differences among methods. Once mixed-event cases are selected, the GNN produces more spatially discriminative exceedance probability fields, whereas kriging-based maps tend to be smoother and more diffuse, in line with the over-conservative behavior seen in the reliability analysis. This strengthens the argument that uncertainty calibration is not an abstract modeling objective; it directly affects the interpretability and actionability of downstream risk-oriented products.

Several limitations should nevertheless be acknowledged. First, the sensor observations used in this study are virtual sensors sampled from gridded Daymet fields, rather than measurements from a physical station network. This design is appropriate for controlled benchmarking and method development, but real deployments introduce additional challenges such as instrument noise, calibration drift, siting effects, and communication outages. Second, the current uncertainty formulation is based on a heteroscedastic likelihood learned by a single model and should therefore be interpreted as conditional predictive uncertainty within the trained model class, rather than a full decomposition of aleatoric and epistemic uncertainty. Third, while the temporal hold-out protocol (train 2020–2023, test 2024) is a strong test of time generalization within the same region, cross-region transferability and robustness to major domain shifts remain open questions.

These limitations also point naturally to future research directions. A next step would be to evaluate the framework against real sensor networks and investigate domain adaptation between gridded products and in situ observations. Another direction is to enrich uncertainty modeling through ensembles or Monte Carlo approximations to capture model-form uncertainty in addition to heteroscedastic predictive variance. On the sensor design side, it would be useful to examine multi-objective placement criteria that explicitly trade off coverage, modal informativeness, and uncertainty reduction, rather than optimizing a single criterion. Finally, extending the framework to jointly reconstruct multiple meteorological variables (e.g., temperature and humidity) could improve applicability for thermal comfort and heat-risk assessment.

## 5. Conclusion

This study presented an uncertainty-aware graph neural network framework for reconstructing daily maximum temperature fields from sparse sensor observations under practical sensor-spacing constraints. Using a strict temporal hold-out design (train: 2020–2023; test: 2024) on a Montreal-area Daymet subset, the proposed approach combined graph-based reconstruction, heteroscedastic predictive uncertainty, and constraint-aware sensor placement analysis. The results showed that the proposed GNN consistently outperformed IDW and ordinary kriging in deterministic reconstruction accuracy (RMSE/MAE) across all tested sensor budgets. Sensor-placement effects were strongest at low budgets and became smaller at higher budgets, while a diminishing-returns regime emerged near 30 sensors under the 4 km spacing constraint. This provides a practically useful indication of a near-saturation sensor density for the study domain. The uncertainty analysis showed that the GNN produced spatially varying uncertainty estimates whose calibration improved with sensor density, and that it achieved a better probabilistic performance (CRPS/NLL and coverage behavior) than kriging in the tested setup. Qualitative heat-event reconstructions and exceedance probability maps further demonstrated that the framework can provide spatially coherent temperature and uncertainty fields for decision-oriented temperature-risk applications. Overall, the proposed framework offers an effective approach for uncertainty-aware temperature field reconstruction from sparse observations, while also enabling evaluation of sensor placement strategies under deployment constraints. Future work should focus on validation with real sensor networks, richer uncertainty modeling, and extension to additional variables and domains.

**Acknowledgment:** This work was supported by the Natural Sciences and Engineering Research Council of Canada (NSERC) [grant numbers: RGPIN 2022-03492 and RGPIN 2026-04810].